\documentclass[12pt]{iopart}
\usepackage[dvips]{graphicx}
\newcommand{\la}{\left\langle}
\newcommand{\ra}{\right\rangle}
\newcommand{\EPL}{{\it Europhys.~Lett.~}}

\newcommand{\MP}{{\it Mol.~Phys.~}}
\newcommand{\JCIS}{{\it J.~Coll.~Int.~Sci.~}}
\newcommand{\EPJ}{{\it Eur.~Phys.~J.~}}

\begin{document}

\title[Crowding of Polymer Coils in Nanoparticle-Polymer Mixtures]
{Crowding of Polymer Coils and Demixing in Nanoparticle-Polymer Mixtures}

\author{Ben Lu and Alan R Denton\footnote[1]{Corresponding author.
Electronic address: {\tt alan.denton@ndsu.edu}}}
\address{Department of Physics, North Dakota State University,
Fargo, ND 58108-6050, USA}

\date{25 April 2011}
\begin{abstract}
The Asakura-Oosawa-Vrij (AOV) model of colloid-polymer mixtures idealizes
nonadsorbing polymers as effective spheres that are fixed in size and
impenetrable to hard particles.  Real polymer coils, however, are
intrinsically polydisperse in size (radius of gyration) and may be
penetrated by smaller particles.  Crowding by nanoparticles can affect
the size distribution of polymer coils, thereby modifying effective
depletion interactions and thermodynamic stability.  To analyse the
influence of crowding on polymer conformations and demixing phase behaviour,
we adapt the AOV model to mixtures of nanoparticles and ideal, penetrable
polymer coils that can vary in size.  We perform Gibbs ensemble Monte Carlo
simulations, including trial nanoparticle-polymer overlaps and variations in
radius of gyration.  Results are compared with predictions of free-volume theory.
Simulation and theory consistently predict that ideal polymers are compressed
by nanoparticles and that compressibility and penetrability stabilise
nanoparticle-polymer mixtures.
\end{abstract}


\maketitle
\newpage


\section{Introduction}
Colloid-polymer mixtures are among the most actively studied materials because of the
windows they open onto the rich physical behaviour of both soft (macromolecular) and
hard (atomically ordered) condensed matter.  The close analogy between colloidal particles
and atoms, combined with real-time imaging of slowly moving particles,
have deepened our fundamental understanding of phase transitions, from freezing/melting
to the glass transition, gelation, and demixing~\cite{pusey91,poon02,tuinier_acis03}.
With steady advances in nanoparticle synthesis and characterization, much recent interest
has turned to the physical properties of nanoparticle-polymer composites~\cite{balazs06}.

Adding nonadsorbing polymers to a stable suspension of colloids or nanoparticles
can induce aggregation and demixing through a depletion mechanism first explained
over a half-century ago by Asakura and Oosawa~\cite{ao54}.  Depletion of polymers
from the space between two particles creates an unbalanced osmotic pressure that
drives the particles together.  Equivalently, the larger volume available to
polymers amidst particles whose excluded-volume shells overlap increases the
polymer entropy and manifests as an effective interparticle attraction.
Tuning the range and strength of the attraction by varying the polymer size and
concentration directly influences phase stability.

The most widely studied model of mixtures of particles and nonadsorbing polymers is
the Asakura-Oosawa-Vrij (AOV) model~\cite{ao54,vrij76}, which treats the particles
as hard spheres and the polymers as mutually non-interacting (ideal) effective spheres
of fixed radius, having hard interactions with the particles.  Despite its simplicity,
this important reference model succeeds in qualitatively explaining demixing,
freezing, and other phenomena observed in real particle-polymer mixtures.
Bulk thermodynamic phase behaviour of the AOV model has been intensively
explored using thermodynamic perturbation theory~\cite{gast86}, free-volume
theory~\cite{lekkerkerker92}, and a variety of computational methods.
Most simulation studies have been based on Monte Carlo algorithms implemented
in either the Gibbs ensemble~\cite{panagiotopoulos00,panagiotopoulos87,
panagiotopoulos88,panagiotopoulos89,panagiotopoulos95} or the grand canonical
ensemble~\cite{vink-horbach_jpcm04,vink-horbach_jcp04,loverso-vink-reatto06},
both of which circumvent complications associated with phase interfaces.
Quantitative discrepancies between predictions and experiments have motivated
enhancements of the AOV model to include more realism in the polymer properties,
including polymer-polymer interactions~\cite{louis00,bolhuis02,schmidt02,schmidt03,
tuinier_pccp03,rosch-errington08,zausch-virnau_jcp09,zausch-horbach_jpcm10},
polydispersity in molecular weight~\cite{sear97,warren97,fasolo05}, and
conformational freedom of polymers on a lattice~\cite{meijer-frenkel91,panagiotopoulos06}.
Recent theoretical and simulation studies of particle-polymer mixtures also have
explored interfacial properties~\cite{moncho-jorda_jcp03,vink-schmidt05,
dijkstra-van-roij-roth-fortini06},
demixing in confinement~\cite{schmidt-fortini-dijkstra03,fortini-schmidt-dijkstra06,
vink-binder-horbach06,vink-devirgilis-horbach06,devirgilis-vink-horbach-binder07,
devirgilis-vink-horbach-binder08}, and dynamical properties, such as diffusion and
response to shear~\cite{zausch-virnau_jcp09,zausch-horbach_jpcm10}.

A polymer coil has a size that is well characterized by its radius of gyration~\cite{degennes}.
Scattering experiments (using, e.g., light or neutrons) probe the mean coil size
averaged over an ensemble of conformations.  Even for a hypothetical solution having
uniform molecular weight (i.e., chain length for linear polymers), the (pre-averaged)
radius of gyration exhibits a broad distribution due to statistical fluctuations
in coil conformations~\cite{fixman62,flory-fisk66,flory69}.  This intrinsic polydispersity
can influence the phase behaviour of particle-polymer mixtures.

To accommodate intrinsic polymer polydispersity in particle-polymer mixtures,
Denton and Schmidt~\cite{denton02} recently proposed a modified AOV model in which
the polymer coils have a single internal degree of freedom, namely size.
Taking as input the radius of gyration distribution of an ideal polymer coil,
they developed and applied a classical density-functional theory, which reduces to
a free-volume theory for uniform fluids.
In the ``colloid" limit, in which the polymer coils are impenetrable to the larger
particles, the theory predicts compression of polymer with increasing particle
concentration, and a resultant stabilisation of the mixture.

In the ``protein" (or nanoparticle) limit, the particles are small enough to penetrate
the polymer coils.  Recent experiments~\cite{caseri00,zukoski-langmuir02,
zukoski-jcp02,zukoski-jcp03,kramer_jcp05,kramer_macromol05,hennequin05},
computer simulations~\cite{panagiotopoulos06,moncho-jorda_jpcm03,depablo04,depablo05},
and theories~\cite{moncho-jorda_jpcm03,schmidt-fuchs02,bolhuis03,belotelov05}
have begun to explore the behaviour of such asymmetric mixtures.  These studies raise
prospects of modifying protein solutions by adding polymer and tuning properties of
polymer-metal (or polymer-semiconductor) nanocomposites by adding nanoparticles
to a polymer matrix.  Depending on solvent quality and other sample conditions,
experiments indicate compression (shrinking)~\cite{huber05,longeville09}, expansion
(swelling)~\cite{mackay08}, or little change in size~\cite{sen07} of polymers
in response to nanoparticles.  Monte Carlo simulations of bead-spring polymers in the
presence of nanoparticles~\cite{depablo04,depablo05} indicate that nanoparticles can
penetrate much larger polymer coils.  Bulk demixing behaviour has been explored via
density-functional theory within a relatively simple model in which the nanoparticles
can penetrate the polymers (effective spheres of fixed size) after surmounting an
energy barrier~\cite{schmidt-fuchs02}.

The main purpose of this paper is to study, by means of simulation and theory,
the influence of nanoparticles on the conformations of nonadsorbing polymers
and the resulting phase stability of nanoparticle-polymer mixtures.
Section~\ref{Model} first defines a simple extension of the AOV model that
incorporates nanoparticle-polymer overlap and intrinsic polydispersity in polymer size.
Section~\ref{Methods} next outlines our methods: Gibbs ensemble Monte Carlo simulation
and a mean-field free-volume theory (with details consigned to appendices).
Section~\ref{Results} presents simulation results and theoretical predictions for
polymer size distributions and demixing phase diagrams.
Finally, Sec.~\ref{Conclusions} summarizes and concludes.

\section{Model}\label{Model}
We consider a mixture of impenetrable nanoparticles and nonadsorbing polymers,
dispersed in a solvent, in osmotic equilibrium with a reservoir of pure polymer solution.
Exchange of polymers between the system and reservoir (e.g., via a semi-permeable membrane)
maintains constant polymer chemical potential.  The thermodynamic state of the system
is characterized by the temperature $T$, the nanoparticle number density $\rho_n$, and
the polymer number density in the reservoir $\rho_p^r$.  Equality of polymer chemical
potentials in the system and reservoir determines the polymer density in the system $\rho_p$.

To model this system, we extend the AOV model to the protein limit in the manner proposed
by Schmidt and Fuchs~\cite{schmidt-fuchs02} by allowing nanoparticles to penetrate polymers
and attributing to each overlapping nanoparticle-polymer pair an energy cost $\epsilon$.
With the fixed nanoparticle radius denoted by $R_n$ (diameter $\sigma_n=2R_n$) and the
instantaneous radius of a polymer by $R_p$, the nanoparticle-nanoparticle and
nanoparticle-polymer interactions are specified by pair potentials:
\begin{eqnarray}
v_{nn}(r)&=&\left\{\begin{array}{l@{\quad\quad}l}
\infty, \qquad & r<\sigma_n~, \\
0, \qquad & r\geq \sigma_n~,
\end{array} \right.
\\
v_{np}(r)&=&\left\{\begin{array}{l@{\quad\quad}l}
\epsilon, \qquad & r<R_n+R_p~, \\
0, \qquad & r\geq R_n+R_p~,
\end{array} \right.
\label{vr}
\end{eqnarray}
where $r$ is the centre-to-centre distance.  Assuming ideal polymers,
the polymer-polymer interaction vanishes, i.e., $v_{pp}(r)=0$ for all $r$, which
strictly applies only to theta solvents~\cite{degennes}, wherein the polymer
second virial coefficient vanishes.

We further extend the AOV model to describe polymer coils whose size distribution is
intrinsically polydisperse~\cite{denton02}.  As noted above, even an idealized
solution of polymers with uniform chain length has a broad distribution of
radius of gyration.  In fact, the radius of gyration of an ideal, freely-jointed
chain follows the exact probability distribution~\cite{fujita70,yamakawa70}
\begin{eqnarray}
P_r(R_p)&=&\frac{1}{\sqrt{2}\pi R_g^r t^3}  \sum_{k=0}^\infty
\frac{(2k+1)!}{(2^{k}k!)^{2}}(4k+3)^{7/2} \exp(-t_k) \nonumber \\[0.5ex]
&\times&\left[\left(1-\frac{5}{8 t_k}\right)K_{1/4}(t_k)+
\left(1-\frac{3}{8 t_k}\right)K_{3/4}(t_k)\right]~,
\label{PR}
\end{eqnarray}
where
\begin{equation}
R_g^r~=~\sqrt{\int_0^{\infty}{\rm d}R_p\, R_p^2\, P_r(R_p)}
\end{equation}
is the root-mean-square radius of gyration of polymers in the reservoir, $t=(R_p/R_g^r)^2$,
$t_k=(4k+3)^2/(8t)$, and $K_n$ are modified Bessel functions of the second kind.

The extended AOV model is fully specified by the polymer size distribution [equation (\ref{PR})]
and two parameters: the penetration energy $\epsilon$ and the ratio of the
rms radius of gyration of polymers in the reservoir to the nanoparticle radius,
$q_r\equiv R_g^r/R_n$, which is an experimentally accessible property.
In the colloid limit ($q_r\leq 1$), we assume impenetrable polymers by taking infinite
penetration energy ($\epsilon\to\infty$) in equation (\ref{vr}).  In the protein limit
($q_r\gg 1$), in which the particles can penetrate the polymers, we follow Schmidt and
Fuchs~\cite{schmidt-fuchs02} and take for the penetration energy an approximation
from polymer field theory~\cite{eisenriegler96,eisenriegler99} for the average
excess free energy cost of inserting a hard sphere into an ideal coil:
\begin{equation}
\beta\epsilon~=~\frac{3}{q}~,
\label{epsilon}
\end{equation}
where $\beta\equiv 1/k_{\rm B}T$ and $q\equiv R_p/R_n$ for a polymer of radius $R_p$.
The loss of conformational entropy of a polymer that harbours a nanoparticle is
thus modelled by an energy penalty that decreases as the polymer swells.
(More refined models replace the step-function profile of equation (\ref{epsilon})
by a continuous function of separation~\cite{forsman08,forsman09}.)
Next we explore the combined influences of penetration and compressibility on
polymer size and demixing.

\section{Methods}\label{Methods}
\subsection{Gibbs Ensemble Monte Carlo Simulation}\label{gemc}
The Gibbs ensemble Monte Carlo (GEMC) method~\cite{panagiotopoulos00,panagiotopoulos87,
panagiotopoulos88,panagiotopoulos89,panagiotopoulos95} provides a computationally efficient
means of computing phase coexistence curves for model fluids.  Allowing each phase to
occupy a separate box avoids any need to simulate interfaces.  This method is known
to be of limited accuracy near critical points~\cite{valleau98,panagiotopoulos98,
errington03,hynninen08}, where grand canonical ensemble methods~\cite{vink-horbach_jpcm04,
vink-horbach_jcp04,loverso-vink-reatto06} can yield higher resolution.  For our purposes
of exploring nanoparticle-polymer demixing, however, GEMC proves practical.

We work in the semigrand ensemble, where temperature, total number of nanoparticles, and
total volume, are fixed, while exchange of polymers with a reservoir fixes the
polymer chemical potential.
The conventional GEMC algorithm for our system involves four types of trial move:
(1) displacements of nanoparticles and polymers within each box to ensure thermal equilibrium;
(2) exchanges of volume between boxes to ensure mechanical equilibrium,
characterized by equality of pressures;
(3) transfers of nanoparticles and polymers between boxes to ensure chemical equilibrium,
characterized by equality of chemical potentials for each species; and
(4) transfers of polymers between each box and the reservoir.
The acceptance probabilities for these standard trial moves are given in Appendix A.

To explore the influence of nanoparticles on the polymer size distribution, we
implement an additional trial move: variation in polymer radius of gyration $R_p$.
This move allows the polymer size distribution to adjust to the presence of particles.
A polymer coil has radius of gyration $R_p$ with probability
$P(R_p)\propto P_r(R_p)\exp(-\beta U)$, where $P_r(R_p)$ is the polymer size distribution
of equation (\ref{PR}) and $U$ is the potential energy due to any penetrating particles.
A trial change in a polymer's radius of gyration, from its old value $R_p^o$ to a new
value $R_p^n$, with an attendant change in potential energy $\Delta U$, is then accepted
with probability (see Appendix A)
\begin{equation}
{\cal P}_{\rm size} = \min\left\{1,~\frac{P_r(R_p^n)}{P_r(R_p^o)}
\exp(-\beta\Delta U)\right\}~.
\label{size-variation}
\end{equation}

Through trial expansions and contractions, the polymers achieve their equilibrium
size distribution.  In a dilute suspension of nanoparticles, the size distribution approaches
that of the polymer reservoir $P_r(R_p)$.  With increasing nanoparticle concentration,
however, crowding and penetration influence the polymer size distribution.
Exploring the shift in polymer size distribution induced by nanoparticles, and the
resulting effect on demixing, is the main goal of our simulation study.

\subsection{Simulation Details}
We simulated mixtures of nanoparticles and polymers confined to a cubic box (or boxes)
with periodic boundary conditions.  In the protein limit, we performed GEMC simulations
in the semigrand ensemble, with fixed total number of nanoparticles $N_n$, total volume $V$,
and polymer reservoir density $\rho_p^r$, for reservoir size ratio $q_r=3$.  The volumes
of the two boxes, initially set equal, were determined by $N_n$ and the nanoparticle
volume fraction $\eta_n=(4\pi/3)\rho_nR_n^3$.  In the colloid limit, we performed
canonical ensemble simulations (one box), with fixed $N_n$, polymer number $N_p$,
and volume, for $q_r=1$ .

The initial configurations were generated by randomly placing particles and polymers
(initially monodisperse) on fcc lattice sites.  At polymer concentrations exceeding
full lattice occupancy, some sites were multiply occupied by polymers.
The initial total number of polymers was chosen to yield a desired effective
polymer volume fraction, defined as $\eta_p=(4\pi/3)\rho_p(R_g^r)^3$ ---
an experimentally controllable quantity.  From these initial states, the simulations
proceeded via the various trial moves outlined in Sec.~\ref{gemc}.
Tolerances were adjusted to yield practical acceptance ratios for each move.

Several diagnostic quantities were calculated to confirm equilibration.
The pressure $p$ in each box was computed from a simple adaptation of the
virial expression for a binary hard-sphere fluid mixture~\cite{hansen90}:
\begin{equation}
\hspace*{-2cm}
\frac{\beta p}{\rho}=1+\frac{2}{3}\pi\rho\la x_n^2 \sigma_n^3
g_{nn}(\sigma_n)+2(1-e^{-\beta\epsilon})x_n\sum_{R_p}x_p(R_n+R_p)^3
g_{np}(R_n+R_p)\ra,
\label{virial}
\end{equation}
where in the given box $\rho$ is the average total number density,
$x_n$ and $x_p$ are the average concentrations of nanoparticles and polymers,
$g_{nn}(\sigma_n)$ and $g_{np}(R_n+R_p)$ are the contact values of
the nanoparticle-nanoparticle and nanoparticle-polymer radial distribution functions,
the summation runs over all polymer radii, and angular brackets represent
an ensemble average over all configurations of nanoparticles and polymers.
In practice, the contact values of the radial distribution functions were computed
by assigning polymer radii to bins of width $(R_n+R_g^r)/50$, accumulating
particles in radial bins of width 5\% of the radius of the central nanoparticle,
and extrapolating to contact.

The chemical potential of the nanoparticles was computed using Widom's test particle
insertion method~\cite{widom63}, applied to trial nanoparticle transfers:
\begin{equation}
\beta\mu_n=-\ln\left\langle\frac{V/\sigma_n^3}{N_n+1}\exp(-\beta\Delta U)
\right\rangle,
\label{mun}
\end{equation}
where $\Delta U$ is the change in potential energy of the box into which a
nanoparticle is inserted and $\langle~\rangle$ represents an ensemble average over
all configurations.  In the semigrand ensemble, the chemical potentials of
polymers of all sizes are imposed by the reservoir.  As a consistency check,
the mean polymer chemical potential (averaged over polymer sizes) also
was computed via Widom's insertion method applied to trial polymer transfers:
\begin{equation}
\beta\mu_p=-\ln\left\langle\frac{V/\sigma_n^3}{N_p+1}\exp(-\beta\Delta U)
\right\rangle,
\label{mup}
\end{equation}
where $\Delta U$ now is the change in potential energy upon insertion of a polymer.
Note that the chemical potentials are defined relative to an arbitrary
reference potential, determined by the choice of volume units (here $\sigma_n^3$).

For comparison with free-volume theory (Sec.~\ref{theory}), we also computed
the effective fraction of the total volume (averaged over polymer sizes) that
was accessible to the polymers, i.e., not excluded by the nanoparticles.
At equilibrium, the chemical potentials of polymer in the system and reservoir
must be equal.  Thus, equating $\mu_p$ [from equation (\ref{mup})] to the
mean chemical potential of (ideal) polymers in the reservoir,
\begin{equation}
\beta\mu_p^r=\ln\left(\rho_p^r\sigma_n^3\right)
=\ln\left(\rho_p\sigma_n^3/\alpha_{\rm eff}\right)~,
\label{mupr}
\end{equation}
the effective polymer free-volume fraction amidst nanoparticles can be expressed as
\begin{equation}
\alpha_{\rm eff}=\rho_p\left\langle\frac{V}{N_p+1}\exp(-\beta\Delta U)\right\rangle~.
\label{alphaeff-sim}
\end{equation}
The effective polymer free-volume fraction for a given box was determined by applying
equation (\ref{alphaeff-sim}) during the GEMC simulation and averaging over all trial
polymer transfers into the box.

\subsection{Free-Volume Theory}\label{theory}
To guide the choice of parameters in our simulations, we generalize the
free-volume theory of Lekkerkerker \etal~\cite{lekkerkerker92} to the case of
compressible, penetrable polymers.  As shown in Appendix B, the Helmholtz
free energy density $f$ of a mixture of nanoparticles of number density $\rho_n$
(volume fraction $\eta_n$) and polymers in osmotic equilibrium with a polymer
reservoir of density $\rho_p^r$ can be expressed (in $k_BT$ units) as
\begin{equation}
f(\eta_n,\rho_p^r) = \rho_n\left[\ln\left(\rho_n\sigma_n^3\right)-1\right]
+ \phi_{hs}(\eta_n) + \rho_p^r\alpha_{\rm eff}(\eta_n)
\left[\ln\left(\rho_p^r\sigma_n^3\right)-1\right]~,
\label{ffv}
\end{equation}
where
\begin{equation}
\alpha_{\rm eff}(\eta_n) \equiv \int_0^{\infty}{\rm d}R_p\,P_r(R_p)\alpha(R_p;\eta_n)
\label{alphaeff}
\end{equation}
is the {\it effective} free-volume fraction of the polymers amidst nanoparticles of
volume fraction $\eta_n$ [{\it cf}.~equation (\ref{alphaeff-sim})], expressed as
a weighted average of the free-volume fraction $\alpha(R_p;\eta_n)$ of polymers
of radius $R_p$.
The first two terms on the right side of equation (\ref{ffv}) are the ideal-gas and
excess free energy densities of the hard-sphere nanoparticles; the third term is the
total free energy density of the polymers.
Note that in this mean-field approximation, which neglects nanoparticle-polymer correlations,
$\alpha(R_p;\eta_n)$ and $\alpha_{\rm eff}(\eta_n)$ are assumed to be
independent of the polymer density.
The corresponding polymer size distribution in the mixture is given by (see Appendix B)
\begin{equation}
P(R_p;\eta_n) = \frac{\alpha(R_p;\eta_n)}{\alpha_{\rm eff}(\eta_n)}P_r(R_p)~.
\label{PRp}
\end{equation}

\section{Results and Discussion}\label{Results}
We initialized the GEMC simulations with $N_n=400$ nanoparticles and $N_p=1000$ polymers
in a volume yielding a nanoparticle volume fraction of $\eta_n=0.03$.  After equilibrating
for 150,000 MC cycles, statistics were accumulated and averaged over the next 150,000 cycles.
Equilibrium was diagnosed via equality of pressures and chemical potentials in coexisting phases.
Several longer runs for larger systems confirmed that equilibrium was attained and that
finite-size effects were negligible.
The main simulation results reported below are coexistence densities of nanoparticle and
polymer species and probability distributions (histograms) of polymer radius.
\begin{figure}
\begin{center}
\includegraphics[width=0.7\columnwidth]{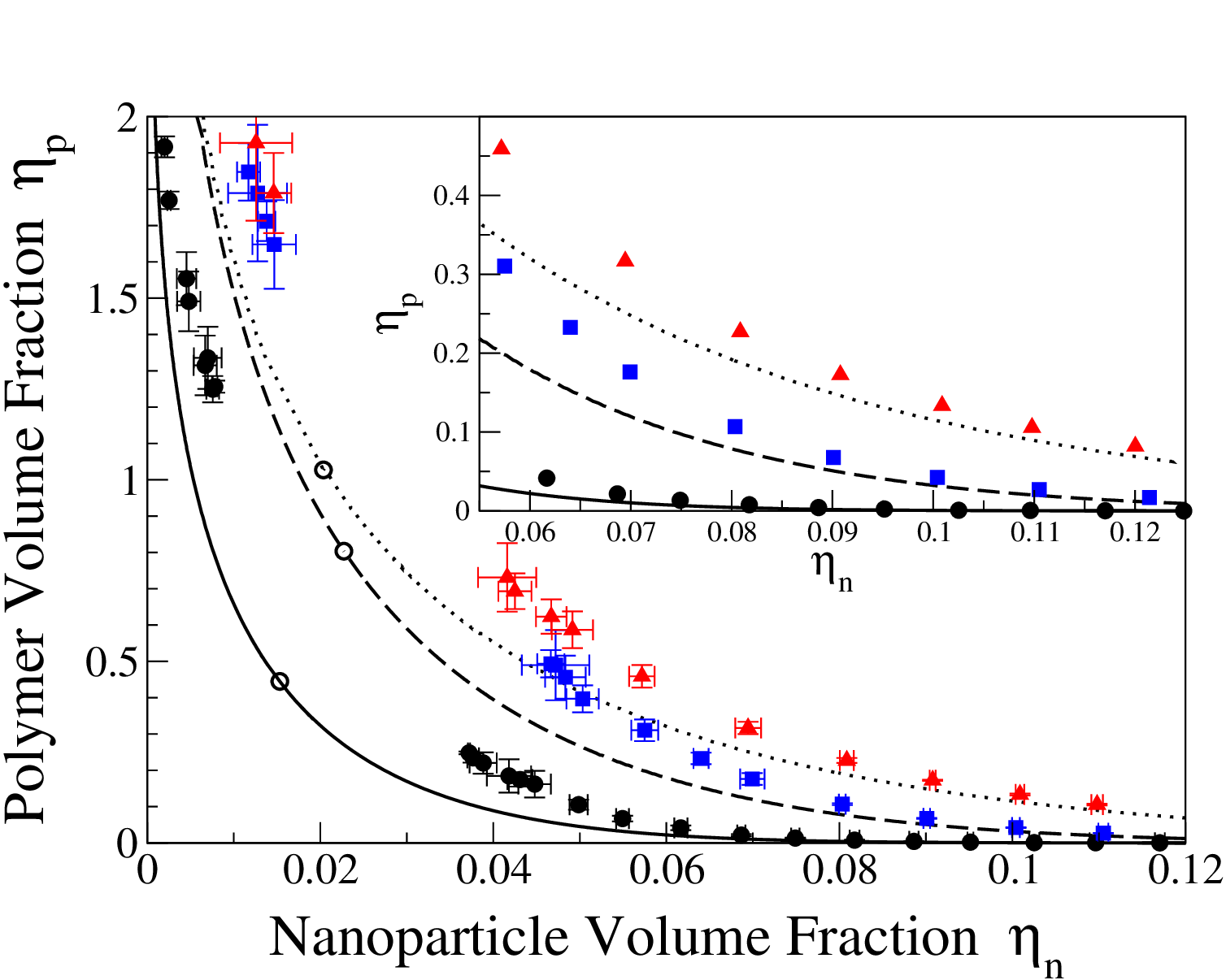}
\end{center}
\caption{\label{fig-pdcp-q3}
Phase diagram of a nanoparticle-polymer mixture in osmotic equilibrium with a
reservoir of polymers whose mean radius of gyration is three times the nanoparticle
radius ($q_r=3$).  Symbols represent simulation data and curves predictions of
free-volume theory for nanoparticle-rich and -poor binodals (open circles are
predicted critical points).  Results are shown for three polymer models:
AOV model with impenetrable, incompressible polymer (black circles, solid curve);
penetrable, incompressible polymer (blue squares, dashed curve); and penetrable,
compressible polymer (red triangles, dotted curve).
Inset: expanded view of nanoparticle-rich phase.
}
\end{figure}

Figure \ref{fig-pdcp-q3} shows the fluid-fluid demixing phase diagram for reservoir
polymer-to-nanoparticle size ratio $q_r=3$ (nanoparticle limit).  In this representation,
the polymer volume fraction in the system (rather than in the reservoir) is plotted on the
vertical axis, facilitating comparison with experiment.  Corresponding points on the
nanoparticle-rich and -poor branches of the binodal represent coexisting ``liquid" and
``vapour" phases.  The demixing and freezing transitions being well separated for this
size ratio, we need not consider the liquid-solid phase boundary.  Results are shown for
the AOV model and its two penetrable-polymer extensions (incompressible and compressible
polymers).  Numerical data from our simulations are plotted together with predictions
from our free-volume theory calculations, derived from a coexistence analysis that equates
pressures and chemical potentials in each phase [using the free energy of equation (\ref{ffv})].

The simulations indicate that polymer penetrability and compressibility both promote stability
against demixing, raising the binodal relative to that of the AOV model.
The free-volume theory predicts the same qualitative trend, but somewhat underestimates
stability of the mixture --- a result of neglecting nanoparticle-polymer correlations.
The theory also predicts an increase in critical polymer concentration with added polymer
freedom.  Although this prediction is impractical to test with the Gibbs ensemble method,
it could be tested by grand canonical ensemble methods.
\begin{figure}
\begin{center}
\includegraphics[width=0.7\columnwidth]{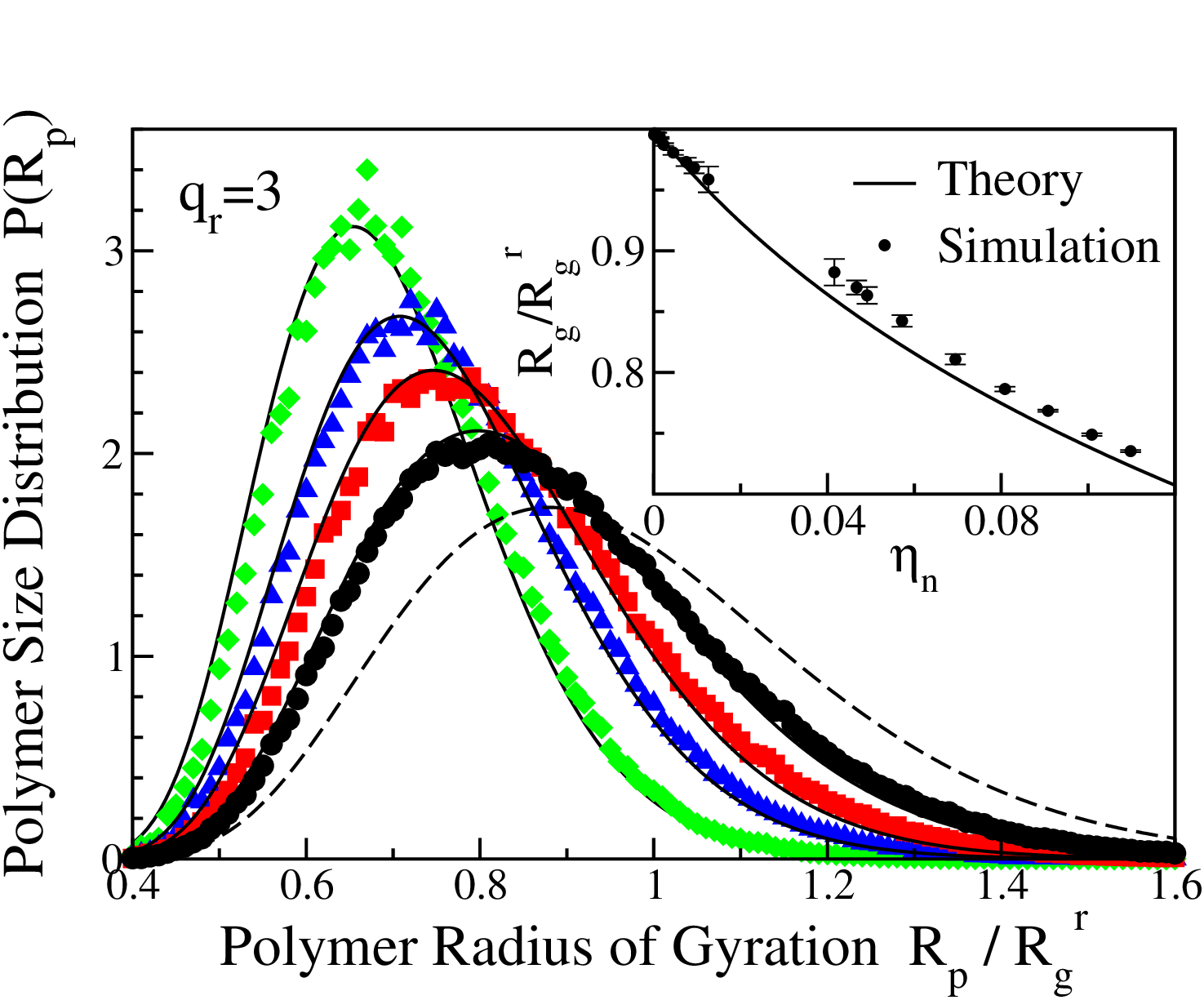}
\end{center}
\caption{\label{fig-pr-q3}
Probability distribution of polymer radius of gyration in a mixture of nanoparticles
and penetrable, compressible polymers in osmotic equilibrium with a polymer reservoir.
The mean radius of polymer in the reservoir is three times the nanoparticle radius
($q_r=3$).  Polymer size distributions are shown along the liquid and vapour binodals
for polymer chemical potentials $-\beta\mu_p=0.9$, 1.3, 1.5, and 1.7 (left to right),
corresponding to mean nanoparticle volume fractions $\eta_n=0.120$, 0.0809, 0.0572,
and 0.0312.  The symbols are simulation data, solid curves are predictions of free-volume
theory, and the dashed curve is the reservoir size distribution (pure polymer solution).
Inset: Ratio of mean polymer radius of gyration in the system $R_g\equiv\la R_p^2\ra^{1/2}$
[equation (\ref{Rg})] to radius of gyration in the reservoir $R_g^r$ (see Table 1).
}
\end{figure}

Polymer radius of gyration distributions (normalized histograms) are shown in
Fig.~\ref{fig-pr-q3} over a range of polymer chemical potentials along the
liquid-vapour binodal.  Our simulation data (smoothed by a running average) are
compared with theoretical predictions from equation (\ref{PRp}).  The size distribution
is seen to shift towards smaller radii of gyration with increasing nanoparticle concentration,
reflecting compression of polymers due to crowding and penetration by nanoparticles.
Endowing each polymer with the internal freedom to change its radius of gyration
allows the polymers to shrink (compress) to avoid penetration by nanoparticles and
contributes to stabilising the mixture.
Polymer compression is further quantified by the mean radius of gyration,
\begin{equation}
R_g(\eta_n)=\la R_p^2\ra^{1/2}=
\left(\Delta R_p\sum_i\,P(R_p^{(i)};\eta_n)(R_p^{(i)})^2\right)^{1/2},
\label{Rg}
\end{equation}
defined as an average over conformations, i.e., sum over histogram bins, where
$R_p^{(i)}$ is the polymer radius at bin $i$, and $\Delta R_p$ is the bin width.
The corresponding polymer compression ratios $R_g/R_g^r$ are plotted in the
inset of Fig.~\ref{fig-pr-q3} and listed in Table~\ref{table1}.
\Table{\label{table1}
Polymer compression ratio $R_g(\eta_n)/R_g^r$ for reservoir polymer-to-nanoparticle
size ratio $q_r=R_g^r/R_n=3$ (see inset to Fig.~\ref{fig-pr-q3}).  Results of
Monte Carlo simulations are compared with predictions of free-volume theory over
a range of nanoparticle volume fractions $\eta_n$ and polymer volume fractions $\eta_p$ along the demixing binodal.} \\
\br
$\eta_n$&$\eta_p$&Simulation&Theory \\
\mr
0.002&3.21&0.987 $\pm$ 0.001&0.991 \\
0.009&2.17&0.968 $\pm$ 0.005&0.963 \\
0.013&1.93&0.959 $\pm$ 0.011&0.948 \\
0.042&0.731&0.882 $\pm$ 0.011&0.859 \\
0.049&0.587&0.863 $\pm$ 0.007&0.841 \\
0.057&0.459&0.842 $\pm$ 0.005&0.822 \\
0.069&0.317&0.811 $\pm$ 0.004&0.796 \\
0.081&0.227&0.786 $\pm$ 0.002&0.773 \\
0.091&0.173&0.768 $\pm$ 0.001&0.754 \\
0.100&0.133&0.749 $\pm$ 0.001&0.739 \\
0.110&0.106&0.735 $\pm$ 0.001&0.723 \\
0.120&0.082&0.720 $\pm$ 0.002&0.707 \\
\br
\end{tabular}
\end{indented}
\end{table}

\begin{figure}
\begin{center}
\vspace*{1cm}
\includegraphics[width=0.7\columnwidth]{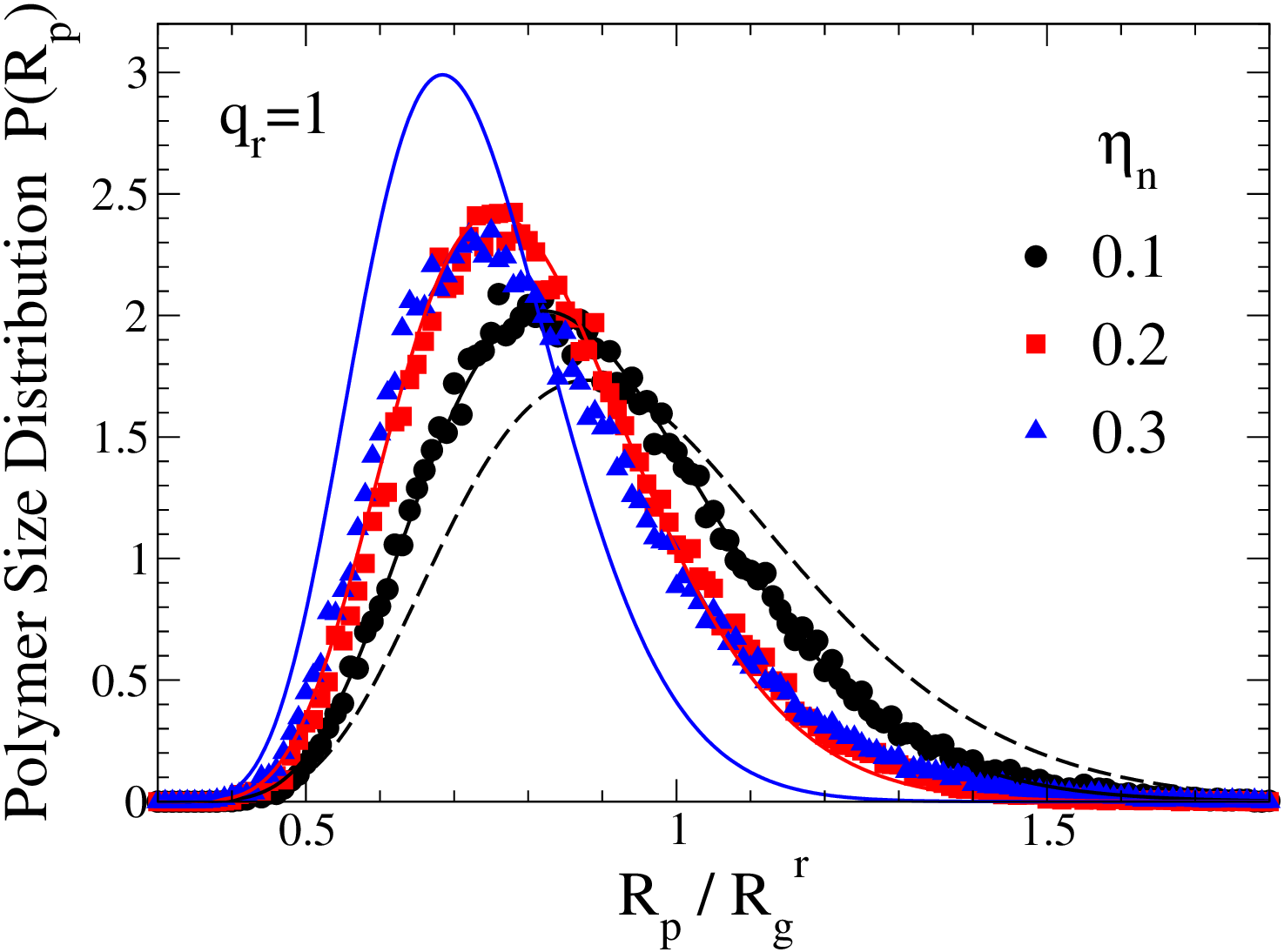}
\end{center}
\vspace*{-0.5cm}
\caption{\label{fig-pr-q1}
Probability distribution of polymer radius of gyration $R_p$ in a mixture of particles
and impenetrable, but compressible, polymer at fixed polymer volume fraction $\eta_p=0.1$.
The mean radius of polymer in the reservoir is equal to the particle radius ($q_r=1$).
Simulation data (symbols) and predictions of free-volume theory (curves) are shown for
increasing particle volume fraction $\eta_n=0$, 0.1, 0.2, and 0.3 (right to left).
}
\end{figure}

To further explore polymer compression, we performed canonical ensemble (constant-$NVT$)
simulations in the colloid limit ($q_r=1$) at fixed polymer volume fraction.  Polymer size
distributions and compression ratios are shown over a range of particle volume fractions
in Fig.~\ref{fig-pr-q1}.  For dilute suspensions, crowding by particles compresses the
polymers, as for $q_r=3$.  These results support the trends in polymer size distribution
predicted previously~\cite{denton02} for the same compressible polymer model.
With increasing particle concentration, however, the polymers cease shrinking and
begin to expand.  This size reversal can be understood by noting that the trajectory
of the state point now crosses the binodal from the stable (mixed) region into the unstable
(demixed) region of the phase diagram.  Snapshots of the system in the unstable region
reveal the presence of large polymer clusters.  Within such a cluster, polymers are
shielded from particles and thus are free to adopt a size distribution closer to that
in the reservoir.  Such correlation-driven behaviour clearly is not captured
by the mean-field theory.

\begin{figure}
\begin{center}
\vspace*{0.5cm}
\includegraphics[width=0.7\columnwidth]{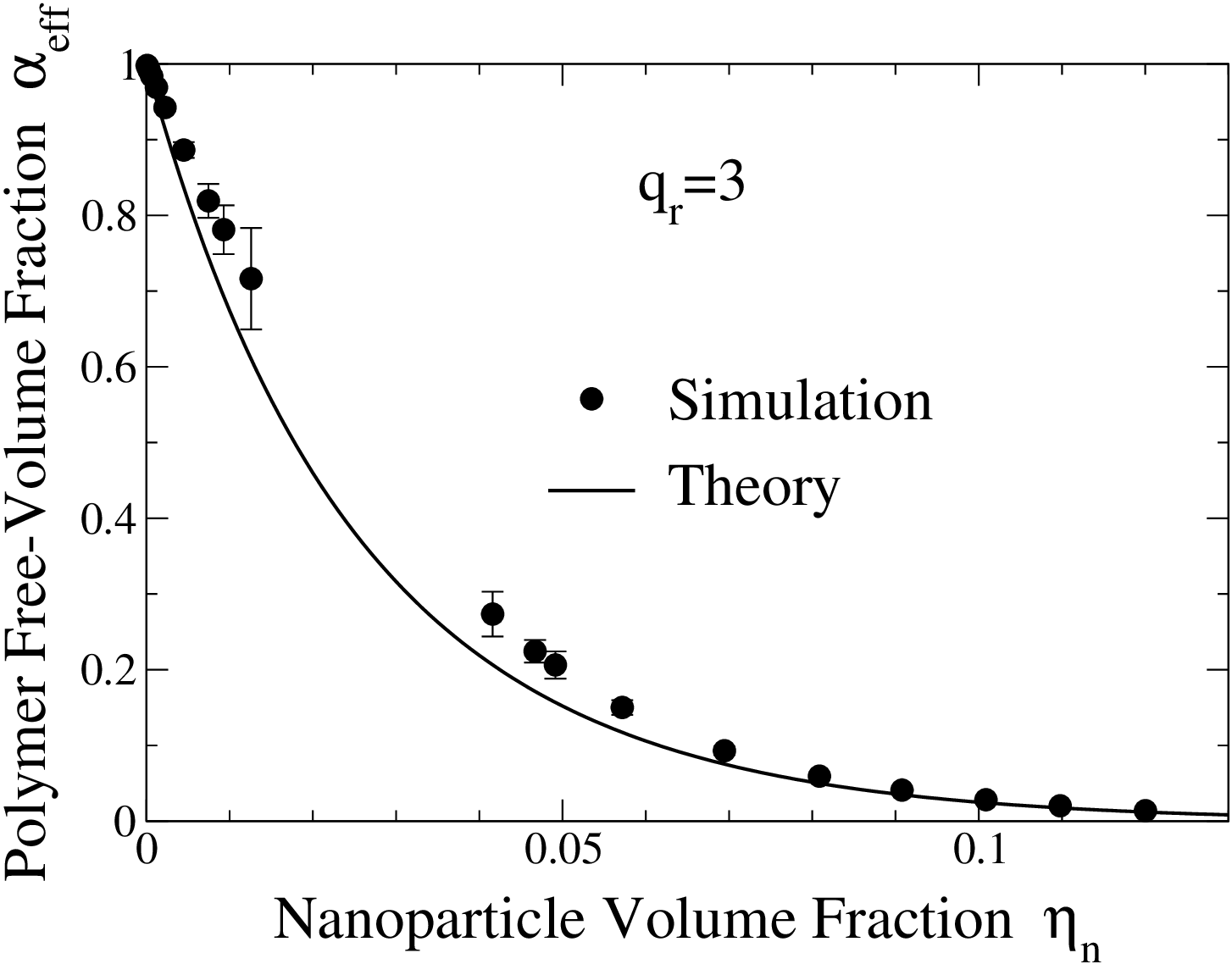}
\end{center}
\vspace*{-0.5cm}
\caption{\label{fig-alphaeff-q3}
Effective polymer free-volume fraction $\alpha_{\rm eff}$ in a mixture of
nanoparticles and penetrable, compressible polymer in osmotic equilibrium with
a polymer reservoir for reservoir polymer-to-nanoparticle size ratio $q_r=3$.
The symbols are simulation data [equation (\ref{alphaeff-sim})] and the curve is
the prediction of free-volume theory [Eq.~(\ref{alphaeff})] along the liquid
and vapour (demixing) binodals.
}
\end{figure}

The influence of polymer compressibility on phase behaviour can be interpreted also in
terms of effective interactions between nanoparticles induced by polymer depletion.
Note that compression of polymers by crowding shortens the range of depletion-induced
attraction between nanoparticles to an extent that depends on particle concentration.
Since polymers in the nanoparticle-rich phase are more compressed, on average, than those
in the nanoparticle-poor phase, weakening of the effective attraction tends to stabilise
the nanoparticle-rich phase against demixing.

Figure \ref{fig-alphaeff-q3} finally presents results for the polymer free-volume fraction
$\alpha$, computed from equations (\ref{alphaeff-sim}) and (\ref{alphaeff}), along the
binodal of Fig.~\ref{fig-pdcp-q3}.
Polymer compression clearly increases the free-volume fraction accessible to the
polymers, consistent with the shift of the nanoparticle-rich binodal towards higher
polymer volume fractions.  The free-volume theory captures the qualitative trend,
but underestimates $\alpha$ at higher nanoparticle concentrations, again because of
neglect of nanoparticle-polymer correlations.

\section{Summary and Conclusions}\label{Conclusions}
In summary, we have performed Monte Carlo simulations of model mixtures of
monodisperse hard-sphere nanoparticles and nonadsorbing, intrinsically polydisperse
polymers in a theta solvent.  The polymers are modeled as effective spheres,
each with a single internal degree of freedom (average size), which allows for
fluctuations in radius of gyration.
In addition to conventional Monte Carlo moves, the polymers undergo trial size
variations, with an acceptance probability dependent on nanoparticle concentration.
Within this coarse-grained model, we have investigated the effect of nanoparticle
crowding on polymer size and the influence of polymer compression and penetrability
on the demixing behaviour of nanoparticle-polymer mixtures.

For particles and polymers of comparable size, our simulations confirm the
trends previously predicted by density-functional theory~\cite{denton02},
namely that particles compress the polymer coils and that mixing stability
is enhanced by polymer compression.  In equilibrium fluid-fluid coexistence,
the polymers fractionate into compressed coils in the particle-rich phase
and expanded coils in the particle-poor phase.  In the protein limit,
in which nanoparticles face an energy barrier for penetrating the larger polymers,
our simulations also show significant polymer compression and enhanced stability,
in qualitative agreement with predictions of free-volume theory.
Incorporating into the model excluded-volume interactions between polymer segments
may help to shed light on the conformations of crowded polymers in good solvents.

While our model includes intrinsic polydispersity of polymers, it neglects
polydispersity in chain length (molecular weight), which can be significant
in real polymer systems~\cite{sear97,warren97,fasolo05}.
Fasolo and Sollich~\cite{fasolo05} have applied free-volume theory to mixtures
of particles and polydisperse polymers, finding that polydispersity in chain length
promotes demixing -- an effect just opposite to that of polymer compressibility.
The relative importance of these two competing effects could be assessed by combining,
in a single model, both radius of gyration and chain length polydispersities.

Finally, we emphasize that the present study demonstrates a general and
practical approach to modelling soft matter systems, in which the microscopic
complexity of the constituent macromolecules is considerably reduced through a
coarse-grained approximation, while physically relevant features are retained
and subsumed into a small number of internal degrees of freedom.  Conceptual
insight provided by this mesoscopic approach may help to guide the design of
experiments and the choice of parameters in studies of more detailed molecular
models~\cite{panagiotopoulos06,depablo04,depablo05}.  A similar analysis of
polymer shape variations induced by particle crowding, and the influence
on demixing behaviour, will be the subject of future work.

\ack
We thank Matthias Schmidt for inspiration
and the Center for Computationally Assisted Science and Technology
at North Dakota State University for computing facilities.
This work was supported by the National Science Foundation under Grant
Nos.~DMR-0204020 and EPS-0132289.
Acknowledgement is made to the Donors of the American Chemical Society Petroleum
Research Fund (PRF 44365-AC7) for partial support of this research.

\appendix
\section{Acceptance Probabilities of Monte Carlo Trial Moves}
\setcounter{section}{1}
The acceptance probabilities for the GEMC trial moves follow directly from
the condition of detailed balance~\cite{frenkel01}, according to which the
average rate of transition from an old state ($o$) to a new state ($n$) must equal,
in equilibrium, the average transition rate from state $n$ to state $o$.  Defining
$P(o)$ and $P(n)$ as the probabilities of finding the system in the states $o$ and $n$,
respectively, and $\pi(o \to n)$ as the transition probability between the states,
detailed balance requires that
\begin{equation}
P(o)\pi(o \to n)=P(n)\pi(n\to o)~.
\label{db}
\end{equation}
Assuming that trial moves $o \to n$ and $n\to o$ are attempted at equal rates,
equation (\ref{db}) implies acceptance probabilities in the ratio
\begin{equation}
\frac{{\rm acc}(o \to n)}{{\rm acc}(n \to o)}=\frac{P(n)}{P(o)}~.
\label{acc}
\end{equation}
In the Metropolis algorithm, trial moves are accepted with probability~\cite{frenkel01,allen}
\begin{equation}
{\cal P}(o \to n) = \min\left\{1,~\frac{P(n)}{P(o)}\right\}~.
\label{Metropolis}
\end{equation}
Trial displacements of randomly chosen particles are thus accepted with probability
\begin{equation}
{\cal P}_{\rm disp} = \min\left\{1,~\exp(-\beta\Delta U )\right\}~,
\label{acc-disp}
\end{equation}
where $\Delta U=U_n-U_o$ is the change in energy between old and new configurations.
A trial move leading to overlap of nanoparticles yields infinite $\Delta U$ and so is
automatically rejected.  Each nanoparticle-polymer overlap increases $U$ by $\epsilon$;
otherwise $\Delta U=0$.

In the semigrand Gibbs ensemble, box $i$ holds $N_i$ particles in a volume $V_i$ ($i=1,2$).
A trial exchange of volume $\Delta V$, achieved by uniformly rescaling all particle
coordinates, such that $V_1\to V_1+\Delta V$ and $V_2\to V_2-\Delta V$,
is accepted with probability
\begin{equation}
{\cal P}_{\rm vol} = \min\left\{1,~\left(\frac{V_1+\Delta V}{V_1}
\right)^{N_1}
\left(\frac{V_2-\Delta V}{V_2}\right)^{N_2} \exp(-\beta\Delta U)\right\}~.
\label{acc-vol1}
\end{equation}
In practice, it proves more efficient to make trial moves in $\ln(V_1/V_2)$,
for which the acceptance probability is~\cite{frenkel01,allen}
\begin{equation}
{\cal P}_{\rm vol} = \min\left\{1,~\left(\frac{V_1+\Delta V}{V_1}
\right)^{N_1+1}
\left(\frac{V_2-\Delta V}{V_2}\right)^{N_2+1}\exp(-\beta\Delta U)\right\}~.
\label{acc-vol2}
\end{equation}

Denoting by $N_{j1}$ and $N_{j2}$ the respective particle numbers of species $j$
($j=n,p$) in the two boxes, transfer of a particle of species $j$ from box 1 to box 2,
resulting in $N_{j1}\to N_{j1}-1$ and $N_{j2}\to N_{j2}+1$, is accepted with
probability~\cite{panagiotopoulos88}
\begin{equation}
{\cal P}_{\rm trans} = \min\left\{1,~\frac{N_{j1}V_2}{(N_{j2}+1)V_1}
\exp(-\beta\Delta U)\right\}.
\label{acc-transfer}
\end{equation}
In the colloid limit ($q_r\le 1$), because of the difficulty of inserting a
large particle into a box without overlaps, only direct transfers of polymers
between the two boxes were attempted.
Transfers of particles were achieved indirectly by exchanging identities
of particles and polymers~\cite{panagiotopoulos89,panagiotopoulos95}.
In an identity-exchange move, a randomly chosen polymer in a randomly
chosen box is changed to a particle; in the other box, a randomly chosen
particle is changed to a polymer.  Changing a polymer to a particle in box 1
(and a particle to a polymer in box 2) is accepted with probability
\begin{equation}
{\cal P}_{\rm ex} = \min\left\{1,~\frac{N_{p1}N_{n2}}{(N_{n1}+1)(N_{p2}+1)}
\exp[-\beta(\Delta U_1+\Delta U_2)]\right\}~,
\label{acc-exchange}
\end{equation}
where $N_{n1}$ ($N_{p1}$) and $N_{n2}$ ($N_{p2}$) are the respective numbers
of particles (polymers) in the two boxes and $\Delta U_1$ and $\Delta U_2$
are the resulting changes in energy in boxes 1 and 2 -- incremented by
$\epsilon$ for each overlap of a particle and a polymer.

Trial transfers of polymer between the system and the reservoir are attempted with
equal frequencies in either direction.  Transfer of a polymer from the reservoir to
the system is executed as follows.  A polymer with a radius chosen randomly from the
reservoir size distribution is placed at a random position in a randomly chosen box.
For a transfer from the reservoir to box 1, the acceptance probability is
\begin{equation}
{\cal P}_{\rm res} = \min\left\{1,~\frac{\rho_p^r V_1}{N_{p1}+1}
\exp(-\beta\Delta U_1)\right\}~,
\label{acc-exchange1}
\end{equation}
where $\Delta U_1$ is the change in energy of box 1.
Transfer of a polymer from the system to the reservoir proceeds by removing
a randomly chosen polymer from a randomly chosen box, with acceptance probability
\begin{equation}
{\cal P}_{\rm res} = \min\left\{1,~\frac{N_{p1}-1}{\rho_p^r V_1}
\exp(-\beta\Delta U_1)\right\}~.
\label{acc-exchange2}
\end{equation}

Finally, from equation (\ref{Metropolis}), a trial change in polymer size
from radius $R_p^o$ to $R_p^n$ is accepted with a probability given by
equation (\ref{size-variation}).
The same result follows also from the general relation
${\cal P}(o \to n)=\min\left\{1,\exp(-\beta\Delta F)\right\}$, where
$\Delta F$ is the change in Helmholtz free energy, if the conformational entropy
of a polymer in the system is taken to be the same as in the reservoir,
i.e., $-k_{\rm B}T\ln P_r(R_p)$.

\section{Free-Volume Theory of Nanoparticle-Polymer Mixtures}
\setcounter{section}{2}
Following \cite{lekkerkerker92,denton02}, the Helmholtz free energy density $f$
of a mixture of nanoparticles of number density $\rho_n$ (volume fraction $\eta_n$)
and polymers in osmotic equilibrium with a polymer reservoir of density $\rho_p^r$
can be expressed (in $k_BT$ units) as
\begin{equation}
f(\eta_n,\rho_p^r) = \rho_n\left[\ln\left(\rho_n\sigma_n^3\right)-1\right]
+ \phi_{hs}(\eta_n) + f_p(\eta_n, \rho_p^r)~,
\label{appendix-ffv}
\end{equation}
to within an arbitrary constant.  The excess free energy density of the
hard-sphere nanoparticles is accurately approximated by the Carnahan-Starling
expression~\cite{hansen90}:
\begin{equation}
\phi_{hs}(\eta_n)=\rho_n\frac{\eta_n(4-3\eta_n)}{(1-\eta_n)^2}~.
\label{appendix-phi}
\end{equation}
Progress in approximating the polymer free energy density $f_p$ is facilitated
by defining polymer density distributions $\rho_p(R_p;\eta_n)$ and $\rho_p^r(R_p)$,
in the system and reservoir, respectively.
These density distributions are related to the corresponding polymer size distributions,
$P(R_p;\eta_n)$ and $P_r(R_p)$, according to
\begin{equation}
\rho_p(R_p;\eta_n)=\rho_p(\eta_n) P(R_p;\eta_n)\quad{\rm and}\quad
\rho_p^r(R_p)=\rho_p^r P_r(R_p)~,
\label{rhopRp}
\end{equation}
which are normalized such that
\begin{equation}
\rho_p(\eta_n)=\int_0^{\infty}{\rm d}R_p\,\rho_p(R_p;\eta_n)\quad{\rm and}\quad
\rho_p^r=\int_0^{\infty}{\rm d}R_p\,\rho_p^r(R_p)
\label{rhop}
\end{equation}
are the average polymer densities in the system and reservoir, respectively.

In free-volume theory, the polymer free energy is approximated by the sum of
the free energy of an ideal gas of polymers confined to the {\it free} volume,
i.e., the volume not excluded by the nanoparticles, and the entropy of a polymer coil
due to internal (conformational) degrees of freedom:
\begin{equation}
f_p = \int_0^{\infty}{\rm d}R_p\,\rho_p(R_p;\eta_n)\left[\ln\left(\frac{\rho_p(R_p;\eta_n)
\sigma_n^3}{\alpha(R_p;\eta_n)}\right)-1+f_{\rm conf}(R_p)\right]~,
\label{fp1}
\end{equation}
where $\alpha(R_p;\eta_n)$ is the free-volume fraction of polymers of radius $R_p$
amidst nanoparticles of volume fraction $\eta_n$ and $f_{\rm conf}(R_p)$ is the
conformational entropy of a polymer coil.
Now equality of polymer chemical potentials in the system and reservoir implies
\begin{equation}
\rho_p(R_p;\eta_n)=\rho_p^r(R_p)\alpha(R_p;\eta_n)~.
\label{alpha}
\end{equation}
Furthermore, assuming the conformational entropy of a polymer in the system
is the same as in the reservoir, we have
\begin{equation}
f_{\rm conf}(R_p) = -\ln P_r(R_p)~.
\label{fconf}
\end{equation}
Combining equations (\ref{rhopRp})-(\ref{fconf}), the polymer free energy density
is given by
\begin{equation}
f_p(\eta_n;\rho_p^r) = \rho_p^r\alpha_{\rm eff}(\eta_n)
\left[\ln\left(\rho_p^r\sigma_n^3\right)-1\right]~,
\label{fp2}
\end{equation}
with the {\it effective} polymer free-volume fraction $\alpha_{\rm eff}(\eta_n)$
defined by equation (\ref{alphaeff}).  Equations (\ref{appendix-ffv}) and (\ref{fp2})
together yield the total free energy density of equation (\ref{ffv}).
Moreover, from equations (\ref{rhopRp}), (\ref{rhop}), and (\ref{alpha}),
the polymer size distribution in the system is expressed by equation (\ref{PRp}).

In practice, it is convenient to approximate the reservoir polymer size distribution
[equation (\ref{PR})] by the accurate, analytic ansatz of Eurich and Maass~\cite{eurich01}:
\begin{equation}
P_r(u) = \frac{1}{2uK_0}\exp\left(-\frac{u}{a}-d^2\frac{a}{u}\right)~,
\label{Eurich-Maass}
\end{equation}
where $u\equiv R_p^2$, $K_0=0.015923$, $a=0.0802$, and $d=1.842$ (see equation (13)
in \cite{eurich01}).  For the polymer free-volume fraction, we adapt the
geometry-based approximation of Oversteegen and Roth~\cite{roth06}:
\begin{equation}
\alpha(R_p;\eta_n) = (1-\eta_n')\exp[-\beta(pv_p+\gamma a_p+\kappa c_p)]~,
\label{alpha-fmt}
\end{equation}
where $\eta_n'\equiv[1-\exp(-\beta\epsilon)]\eta_n$ represents an effective
nanoparticle volume fraction for penetrable polymers~\cite{schmidt-fuchs02}.
Equation (\ref{alpha-fmt}),
a generalization of scaled-particle theory derived from the fundamental measures
formulation of density-functional theory~\cite{rosenfeld89,rosenfeld94,schmidt00}
conceptually separates thermodynamic properties of the nanoparticle hard-sphere fluid ---
bulk pressure $p$, surface tension at a planar hard wall $\gamma$, and bending rigidity
$\kappa$ --- from geometric properties of the polymer depletant ---
volume $v_p=(4\pi/3)R_p^3=(4\pi/3)R_n^3q^3$, surface area $a_p=4\pi R_p^2=4\pi R_n^2q^2$,
and integrated mean curvature $c_p=R_p=R_nq$.  The hard-sphere properties are
approximated by the Carnahan-Starling expressions~\cite{roth06}:
\begin{eqnarray}
\beta p&=&\frac{3\eta_n'}{4\pi R_n^3}\frac{1+\eta_n'+\eta_n'^2-\eta_n'^3}{(1-\eta_n')^3}
\nonumber \\[0.5ex]
\beta\gamma&=&\frac{3}{4\pi R_n^2}\left[\frac{\eta_n'(2-\eta_n')}{(1-\eta_n')^2}
+\ln(1-\eta_n')\right] \nonumber \\[0.5ex]
\beta\kappa&=&\frac{3\eta_n'}{R_n(1-\eta_n')}~.
\label{CSthermo}
\end{eqnarray}

%













\end{document}